\documentclass[twocolumn,showkeys]{revtex4}
\usepackage[utf8]{inputenc}
\usepackage{graphicx}

\begin{document}

\title{Anisotropic pressure effects in hydrodynamic description of waves propagating parallel to the magnetic field in relativistically hot plasmas}

\author{Pavel A. Andreev}
\email{andreevpa@physics.msu.ru}
\affiliation{Department of General Physics, Faculty of physics, Lomonosov Moscow State University, Moscow, Russian Federation, 119991.}

\date{\today}

\begin{abstract}
The structure of novel hydrodynamic model of plasmas with the relativistic temperatures
consisted of four equations for the material fields is presented for the regime of anisotropic pressure and other tensors describing the thermal effects.
Presented model constructed of equation for evolution of the concentration, the velocity field,
the average reverse relativistic $\gamma$ functor, and the flux of the reverse relativistic $\gamma$ functor,
which are considered as main hydrodynamic variables.
Four pressure-like tensors (two second rank tensors and one fourth rank tensor) describe the thermal effects.
Among them we have the flux of the particle current
and the current of the flux of the reverse relativistic $\gamma$ functor.
The high-frequency excitations are considered analytically in order to trace the contribution of the anisotropy of pressure-like tensors in their spectra.
\end{abstract}

\keywords{relativistic plasmas, hydrodynamics, microscopic model, arbitrary temperatures.}


\maketitle





\section{Introduction}

The relativistic and nonrelativistic magnetized plasmas are under active theoretical and experimental study over several decades
due to their important role in the variety phenomena existing in nature.
Particularly, the astrophysical objects show large number of scenarios for the hot plasmas being in the strong magnetic field.
The magnetic field creates the anisotropy in plasmas,
which reveals in the anisotropic structure of tensors describing plasma dynamics at the hydrodynamic description of plasmas
and anisotropy of the distribution functions at the kinetic description of plasmas
(see for instance \cite{Hazeltine APJ 2002} and \cite{Mahajan PoP 2002}).
Variety of physical scenarios for the relativistic plasmas are recently considered in literature
\cite{Mahajan PoP 2011}
\cite{Osmanov Un 21},
\cite{Shatashvili PoP 20},
\cite{Soto-Chavez PRE 10},
\cite{Liu PPCF 21},
\cite{Comisso PRL 14},
\cite{Heyvaerts AA 12},
\cite{Munoz EPS 06},
\cite{Brunetti MNRAS 04},
\cite{She 21}.
Moreover, the plasmas in curved spacetime are also under consideration
\cite{Bhattacharjee PRD 19},
\cite{Comisso 19},
\cite{Asenjo 19},
\cite{Darbha 21},
\cite{Chabanov 21}.
Majority of these scenarios involve the strong magnetic field
which creates the anisotropy of the system.


In hydrodynamics the anisotropy reveals itself via the structure of pressure tensor $\hat{P}$.
The pressure tensor $\hat{P}$ is the symmetric second rank tensor.
It can be presented via three independent functions
if coordinate axis are chosen along the main axis of the system
\begin{equation}\label{RHD2021ClLM} \hat{P}=\left(
                                              \begin{array}{ccc}
                                                P_{xx} & 0 & 0 \\
                                                0 & P_{yy} & 0 \\
                                                0 & 0 & P_{zz} \\
                                              \end{array}
                                            \right).
\end{equation}
Otherwise the pressure is presented within six functions.

However, the symmetry of the system can decrease the number of independent functions describing the pressure tensor.
For instance, the presence of the uniform magnetic field in infinite plasmas gives the axial symmetry for the system.
Hence, the pressure tensor is presented within two functions
\begin{equation}\label{RHD2021ClLM} \hat{P}=\left(
                                              \begin{array}{ccc}
                                                P_{\perp} & 0 & 0 \\
                                                0 & P_{\perp} & 0 \\
                                                0 & 0 & P_{\parallel} \\
                                              \end{array}
                                            \right). \end{equation}
If the magnetic field is relatively weak
we can neglect the difference of the pressures $P_{\perp}\approx P_{\parallel}=p$.
Hence the pressure is proportional to the unit matrix $P^{ab}=P\cdot\delta^{ab}$,
where $\delta^{ab}$ is the Kronecker symbol.

Here we estimate the role of the anisotropy of the pressure (and the pressure-like tensors) of electrons
in the properties of waves propagating parallel to the external magnetic filed in the relativistically hot plasmas.

Above we present the discussion of the pressure tensor,
which is the flux of momentum.
However, the model under development does not include the pressure itself.
While two other hydrodynamic functions with the similar physical meaning exist in this model.
Let us describe the structure of the suggested model.
In our description we follow works
\cite{Andreev 2021 05}, \cite{Andreev 2021 06}, \cite{Andreev 2021 07}, \cite{Andreev 2021 08}, \cite{Andreev 2021 09}.
The model includes the concentration of particles $n$ and equation for its evolution,
which has form of the continuity equation.
The continuity equation includes the velocity field $\textbf{v}$ defined via the current of particles $\textbf{j}=n\textbf{v}$.
Therefore, the second equation of this system is the velocity field evolution equation.
The velocity field evolves under the influence of the electromagnetic field presented by vector fields $\textbf{E}$ and $\textbf{B}$.
The terms describing the interaction are considered in the mean-field approximation (the self-consistent field approximation).
Consequently, it contains three following functions
the average reverse gamma factor $\Gamma$,
the average flux of reverse gamma factor $t^{a}$,
and
the second rank tensor of the current of the flux of reverse gamma factor $t^{ab}$. 
The divergence of the flux of the current of particles $p^{ab}$ gives the kinematic mechanism of the evolution of the velocity field as well.
The velocity field evolution equation shows
that the model includes the following low rank tensors:
the scalar function of the average reverse gamma factor $\Gamma$,
and
the vector function of the average flux of reverse gamma factor $t^{a}$.
Therefore, the evolution equations for these functions are included in the model.
The interaction in this equations is also considered in the mean-field approximation.
the evolution of scalar function $\Gamma$ is expressed via the divergence of vector $t^{a}$.
Moreover, the interaction contribution is expressed via the concentration, the velocity field,
and the flux of current of particles $p^{ab}$.
Hence, no additional functions appear there.
The evolution of vector $t^{a}$ is expressed via the divergence of tensor $t^{ab}$.
The interaction contribution in this equation is presented via the concentration $n$,
current of particles $\textbf{j}=n\textbf{v}$, the flux of current of particles $p^{ab}$,
and the fourth rank tensor $M^{abcd}$,
which is the flux of flux of tensor $p^{ab}$.
Overall, we see the contribution of three high-rank tensors
$p^{ab}$, $t^{ab}$, and $M^{abcd}$,
which requires some equations of state.
Model presented in Refs.
\cite{Andreev 2021 05}, \cite{Andreev 2021 06}, \cite{Andreev 2021 07}, \cite{Andreev 2021 08}, \cite{Andreev 2021 09}
applies the equations of state appearing from relations of tensors $p^{ab}$, $t^{ab}$, $M^{abcd}$
with other hydrodynamic functions obtained for the equilibrium state.
These relations are found via the application of the relativistic Maxwell distribution function.
In earlier papers we focus on approximately isotropic form of the second rank tensors.
To this end, in those papers the equations of state are calculated using isotropic form of the relativistic Maxwellian distribution function.

There are examples of generalization of the relativistic Maxwellian distribution function
\cite{Hazeltine APJ 2002},
where the prefactor in front of the exponent contain combination of parameters giving the anisotropy.
Particularly, the parameter proportional to difference of parallel and perpendicular pressures is introduced.
Some methods introduce the two-temperature description of each species,
where one temperature related to the motion parallel to the magnetic field
and the second temperature is related to the motion perpendicular to the magnetic field.
However, the temperature is the scalar function.
Therefore, the effects of anisotropy cannot lead to several temperatures for the species.
Each species has single temperature.
While, it is well-known that the temperatures of different species (the electrons and ions, for example) can be different.

Below we present a hydrodynamic model which is derived from the microscopic model \cite{Andreev 2021 05}, \cite{Andreev 2021 09}.
The microscopic motion of classic relativistic particles is traced to obtain corresponding definitions and equations for the macroscopic functions.
Most simple definition allowing to show the structure of all hydrodynamic functions is the concentration of particles
$n(\textbf{r},t)$,
which is defined in the arbitrary inertial frame \cite{Kuz'menkov 91}, \cite{Drofa TMP 96}, \cite{Andreev PIERS 2012}
\begin{equation}\label{RHD2021ClLM concentration definition} n(\textbf{r},t)=\frac{1}{\Delta}\int_{\Delta}d\mbox{\boldmath $\xi$}\sum_{i=1}^{N}\delta(\textbf{r}+\mbox{\boldmath $\xi$}-\textbf{r}_{i}(t)). \end{equation}
The integral operator counts the number of particles in the vicinity of the point of space.
hence, we have the number of particles in the volume $\Delta$ around point of space $\textbf{r}$ in an arbitrary moment in time $t$.
The subindex $i$ refers to the number of particle, while $N$ is the total number of particles in the system.
Vector $\textbf{r}_{i}(t)$ presents the radius-vector of $i$-th particle.
Vector $\mbox{\boldmath $\xi$}$ presented under integral in (\ref{RHD2021ClLM concentration definition}) scans the $\Delta$-vicinity.

The number and form of other hydrodynamic functions are obtained during the derivation.
each step of derivation shows
which functions can be applied for the extension of the model.
Their definitions appear via operator
\begin{equation}\label{RHD2021ClLM formula for average}\langle ...\rangle\equiv\frac{1}{\Delta}\int_{\Delta}d\mbox{\boldmath $\xi$}
\sum_{i=1}^{N} ... \delta(\textbf{r}+\mbox{\boldmath $\xi$}-\textbf{r}_{i}(t)),\end{equation}
which is illustrated for the concentration above \cite{Andreev 2021 05}.
Therefore, we have the following set of functions:
the current of particles $\textbf{j}=\langle \textbf{v}_{i}(t)\rangle$,
which allows to introduce the velocity field $\textbf{v}=\textbf{j}/n$,
the reverse gamma factor $\Gamma=\langle \frac{1}{\gamma_{i}}\rangle$,
the flux of reverse gamma factor $t^{a}=\langle \frac{1}{\gamma_{i}}v_{i}^{a} \rangle -\Gamma v^{a}$,
the flux of the current of particles $p^{ab}=\langle v_{i}^{a}v_{i}^{b} \rangle-n v^{a}v^{b}$,
the current of the flux reverse gamma factor $t^{ab}=\langle \frac{1}{\gamma_{i}}v_{i}^{a}v_{i}^{b} \rangle-\Gamma v^{a}v^{b}-t^{a}v^{b}- v^{a}t^{b}$,
here $\gamma_{i}=1/\sqrt{1-\textbf{v}_{i}(t)^{2}/c^{2}}$ is the relativistic gamma factor if $i$-th particle.
The energy-momentum density is not presented here since it is included in the hydrodynamic model.
Quantum modernization of this method is developed in literature
\cite{Maksimov QHM 99}, \cite{Andreev PoF 21}, \cite{Andreev JPP 21}, \cite{Andreev Ch 21}.


This paper is organized as follows.
In Sec. II the relativistic hydrodynamic equations are presented and discussed for the isotropic plasmas.
In Sec. III the contribution of the anisotropy in the spectra of collective excitations is considered analytically.
In Sec. IV a brief summary of obtained results is presented.


\section{Relativistic hydrodynamic model}

Here we follow Refs. \cite{Andreev 2021 05}, \cite{Andreev 2021 09},
and generalize isotropic pressure regimes applied in Refs. \cite{Andreev 2021 06}, \cite{Andreev 2021 07}, \cite{Andreev 2021 08}
where a set of relativistic hydrodynamic equations
is obtained and applied for plasmas with the relativistic temperatures.
The model is composed of four equations.
There are other hydrodynamic models of high-temperature relativistic plasmas,
where the interaction of particles is considered in terms of the momentum evolution equation
\cite{Hazeltine APJ 2002}, \cite{Mahajan PoP 2002}, \cite{Shatashvili PoP 20}, \cite{Shatashvili ASS 97}, \cite{Shatashvili PoP 99}, \cite{Mahajan PRL 03}.

First equation in the presented model is the continuity equation \cite{Andreev 2021 05}
\begin{equation}\label{RHD2021ClLM cont via v} \partial_{t}n+\nabla\cdot(n\textbf{v})=0.\end{equation}

Next, the velocity field evolution equation is \cite{Andreev 2021 05}, \cite{Andreev 2021 09}
$$n\partial_{t}v^{a}+n(\textbf{v}\cdot\nabla)v^{a}
+\frac{1}{m}\partial_{b}[p_{\parallel}e_{B}^{a}e_{B}^{b}+p_{\perp}(\delta^{ab}-e_{B}^{a}e_{B}^{b})]$$
$$=\frac{e}{m}\Gamma E^{a}+\frac{e}{mc}\varepsilon^{abc}(\Gamma v_{b}+t_{b})B_{c}
-\frac{e}{mc^{2}}(\Gamma v^{a} v^{b}+v^{a}t^{b}+v^{b}t^{a})E_{b}$$
\begin{equation}\label{RHD2021ClLM Euler for v}
-\frac{e}{mc^{2}}(\tilde{t}_{\parallel}e_{B}^{a}e_{B}^{b}+\tilde{t}_{\perp}(\delta^{ab}-e_{B}^{a}e_{B}^{b}))E_{b}, \end{equation}
where tensor
$p^{ab}=p_{\parallel}e_{B}^{a}e_{B}^{b}+p_{\perp}(\delta^{ab}-e_{B}^{a}e_{B}^{b})$
is the flux of the thermal velocities,
and tensor $t^{ab}=\tilde{t}_{\parallel}e_{B}^{a}e_{B}^{b}+\tilde{t}_{\perp}(\delta^{ab}-e_{B}^{a}e_{B}^{b})$
is the flux of the average reverse gamma-factor.
Here we use unit vector in the direction of the external magnetic field $\textbf{e}_{B}=\textbf{B}_{ext}/B_{ext}=\textbf{e}_{z}$
which coincides with the unit vector of z-axis.
Therefore, if $a=x$ we find $\partial_{b}p^{ab}=\partial_{x}p_{\perp}$ (same for $a=y$).
Similarly, if $a=z$ we find $\partial_{b}p^{ab}=\partial_{z}p_{\parallel}$.
Parameters $m$ and $e$ are the mass and charge of particle,
$c$ is the speed of light,
$\delta^{ab}$ is the three-dimensional Kronecker symbol,
$\varepsilon^{abc}$ is the three-dimensional Levi-Civita symbol.
In equation (\ref{RHD2021ClLM Euler for v}) and below we assume the summation on the repeating index
$v^{b}_{s}E_{b}=\sum_{b=x,y,z}v^{b}_{s}E_{b}$.
Moreover, the metric tensor has diagonal form corresponding to the Minkovskii space,
it has the following sings $g^{\alpha\beta}=\{-1, +1, +1, +1\}$.
Hence, we can change covariant and contrvariant indexes for the three-vector indexes: $v^{b}_{s}=v_{b,s}$.
The Latin indexes like $a$, $b$, $c$ etc describe the three-vectors,
while the Greek indexes are deposited for the four-vector notations.
The Latin indexes can refer to the species $s=e$ for electrons or $s=i$ for ions.
The Latin indexes can refer to the number of particle $j$ at the microscopic description.
However, the indexes related to coordinates are chosen from the beginning of the alphabet,
while other indexes are chosen in accordance with their physical meaning.

The equation of evolution of the averaged reverse relativistic gamma factor includes the action of the electric field
$$\partial_{t}\Gamma+\partial_{b}(\Gamma v^{b}+t^{b})$$
$$=-\frac{e}{mc^{2}}n\textbf{v}\cdot\textbf{E}
\biggl(1-\frac{1}{c^{2}}\biggl(\textbf{v}^{2}+\frac{(p_{\parallel}+2p_{\perp})}{n}\biggr)\biggr)$$
\begin{equation}\label{RHD2021ClLM eq for Gamma}
+2\frac{e}{mc^{4}}nv_{b}E_{a}(p_{\parallel}e_{B}^{a}e_{B}^{b}+p_{\perp}(\delta^{ab}-e_{B}^{a}e_{B}^{b})) .\end{equation}
Function $\Gamma$ is also called the hydrodynamic Gamma function \cite{Andreev 2021 05}.

The fourth and final equation in this set of hydrodynamic equations is the equation of evolution for the thermal part of
current of the reverse relativistic gamma factor (the hydrodynamic Theta function):
\begin{widetext}$$(\partial_{t}+\textbf{v}\cdot\nabla)t^{a}
+\partial_{b}[\tilde{t}_{\parallel}e_{B}^{a}e_{B}^{b}+\tilde{t}_{\perp}(\delta^{ab}-e_{B}^{a}e_{B}^{b})]
+(\textbf{t}\cdot\nabla) v^{a}+t^{a} (\nabla\cdot \textbf{v})
+\Gamma(\partial_{t}+\textbf{v}\cdot\nabla)v^{a}$$
$$
=\frac{e}{m}nE^{a}\biggl[1-\frac{\textbf{v}^{2}}{c^{2}}-\frac{(p_{\parallel}+2p_{\perp})}{nc^{2}}\biggr]
+\frac{e}{mc}\varepsilon^{abc}nv_{b}B_{c}
\biggl[1-\frac{\textbf{v}^{2}}{c^{2}}-\frac{(p_{\parallel}+2p_{\perp})}{nc^{2}}\biggr]$$
$$-\frac{2e}{mc^{3}}\varepsilon^{abc}v^{d}B_{c}
[p_{\parallel}e_{B,b}e_{B,d}+p_{\perp}(\delta_{bd}-e_{B,b}e_{B,d})]
-\frac{2e}{mc^{2}}E_{b}
\biggl(nv^{a}v^{b}
+[p_{\parallel}e_{B}^{a}e_{B}^{b}+p_{\perp}(\delta^{ab}-e_{B}^{a}e_{B}^{b})]\biggr)$$
$$
+\frac{2e}{mc^{4}}nv^{a}v^{b}E_{b}\biggl[\textbf{v}^{2}+\frac{p_{\parallel}+2p_{\perp}}{n}\biggr]
+\frac{4e}{mc^{4}}E_{b} v^{a}v_{c}    [p_{\parallel}e_{B}^{b}e_{B}^{c}+p_{\perp}(\delta^{bc}-e_{B}^{b}e_{B}^{c})]$$
$$+\frac{4e}{mc^{4}}E^{b} v_{b}v_{c}  [p_{\parallel}e_{B}^{a}e_{B}^{c}+p_{\perp}(\delta^{ac}-e_{B}^{a}e_{B}^{c})]
+\frac{2e}{mc^{4}}E_{b} \textbf{v}^{2} [p_{\parallel}e_{B}^{a}e_{B}^{b}+p_{\perp}(\delta^{ab}-e_{B}^{a}e_{B}^{b})]$$
\begin{equation}\label{RHD2021ClLM eq for t a}
+\frac{2e}{mc^{4}}
(M_{\parallel}+2M_{\parallel,\perp})e_{B}^{a}(\textbf{e}_{B}\cdot \textbf{E})
+\frac{2e}{mc^{4}}
(M_{\parallel,\perp}+4M_{\perp})(\delta^{ab}-e_{B}^{a}e_{B}^{b})] E_{b}.\end{equation}
All hydrodynamic equations are obtained in the mean-field approximation (the self-consistent field approximation).
The fourth rank tensor $M^{abcd}$ entering the equation for evolution of the flux of reverse gamma factor
via its partial trace $M^{abcc}=M_{c}^{cab}$.
If we neglect the anisotropy
we construct tensor $M^{abcd}$ of the Kronecker symbols
$M^{abcd}=(M_{0}/3)(\delta^{ab}\delta^{cd}+\delta^{ac}\delta^{bd}+\delta^{ad}\delta^{bc})$.
It gives $M^{xxxx}=M^{yyyy}=M^{zzzz}=M_{0}$.
If we have two pairs of different projections
we obtain $M^{xxyy}=M^{xxzz}=M_{0}/3$.
Otherwise the element of tensor $M^{abcd}$ is equal to zero.
So, for the partial trace $M_{c}^{cab}$ we find $M_{c}^{cab}=(5M_{0}/3)\delta^{ab}$.
Let us go back to the anisotropic case with the anisotropy axis directed parallel to the z-direction.
So, we have $M^{xxxx}=M^{yyyy}\neq M^{zzzz}$.
Moreover, we have modification of elements with two pairs of different projections:
$M^{xxyy}\neq M^{xxzz}=M^{yyzz}$.
Hence, we have four different values of elements of tensor $M^{abcd}$
instead of two values existing in the isotropic case.
For the purpose of small amplitude waves in the linear approximation on the wave amplitudes
we need equations for $t^{x}$ and $t^{y}$.
These equations contain $M_{c}^{cxb}$ and $M_{c}^{cyb}$
which have the following form.
Including the symmetric properties of the fourth rank tensor $M^{abcd}$
we obtain its following representation
$M^{abcd}=M_{\parallel}e_{B}^{a}e_{B}^{b}e_{B}^{c}e_{B}^{d}$
$+M_{\parallel,\perp}
[e_{B}^{a} e_{B}^{b} (\delta^{cd}-e_{B}^{c}e_{B}^{d})
+e_{B}^{a} e_{B}^{c} (\delta^{bd}-e_{B}^{b}e_{B}^{d})
+e_{B}^{a} e_{B}^{d} (\delta^{bc}-e_{B}^{b}e_{B}^{c})
+e_{B}^{b} e_{B}^{c} (\delta^{ad}-e_{B}^{a}e_{B}^{d})
+e_{B}^{b} e_{B}^{d} (\delta^{ac}-e_{B}^{a}e_{B}^{c})
+e_{B}^{c} e_{B}^{d} (\delta^{ab}-e_{B}^{a}e_{B}^{b})]$
$+M_{\perp}[(\delta^{ab}-e_{B}^{a}e_{B}^{b})(\delta^{cd}-e_{B}^{c}e_{B}^{d})
+(\delta^{ac}-e_{B}^{a}e_{B}^{c})(\delta^{bd}-e_{B}^{b}e_{B}^{d})
+(\delta^{ad}-e_{B}^{a}e_{B}^{d})(\delta^{bc}-e_{B}^{b}e_{B}^{c})]$.
Moreover, we keep the relation between $M^{xxxx}$ and $M^{xxyy}$
existing in the isotropic case $M^{xxxx}=3M^{xxyy}$ and $M^{xxyy}\equiv M_{\perp}$.
Since we have inplane isotropy in the plane perpendicular to the anisotropy axis.
Consequently, we find
$M^{abcc}=M_{\parallel}e_{B}^{a}e_{B}^{b}
+M_{\parallel,\perp} (\delta^{ab}+e_{B}^{a}e_{B}^{b})
+M_{\perp}[4(\delta^{ab}-e_{B}^{a}e_{B}^{b}) ]$,
where $e_{B}^{c}e_{B}^{c}=e_{B,c}e_{B}^{c}=1$,
and $(\delta^{cc}-e_{B}^{c}e_{B}^{c})=2$.
Therefore, we obtain $M_{c}^{cxb}\delta E_{b}=(4M_{\perp} +M_{\parallel,\perp}) \delta E_{x}$
and $M_{c}^{cyb}\delta E_{b}=(4M_{\perp} +M_{\parallel,\perp})\delta E_{y}$.
Let us compare this result with the symmetric case.
It would give $M_{\parallel,\perp}=M_{\perp}=M_{0}/3$.
Hence, we find $4M_{\perp} +M_{\parallel,\perp}=5M_{0}/3$
in accordance with equations presented for the symmetric cases described in
Refs. \cite{Andreev 2021 05}, \cite{Andreev 2021 06}, \cite{Andreev 2021 07}, \cite{Andreev 2021 08}, \cite{Andreev 2021 09}.
Let us point out the physical meaning of fourth rank tensor $M^{abcd}$.
Basically, it is the current of flux of $p^{ab}$.
For the explicit definition of tensor $M^{abcd}$ see equation (17) of Ref. \cite{Andreev 2021 05}.
\end{widetext}

The equations of electromagnetic field have the traditional form
presented in the three-dimensional notations
\begin{equation}\label{RHD2021ClLM div B} \nabla \cdot\textbf{B}=0,\end{equation}
\begin{equation}\label{RHD2021ClLM rot E} \nabla\times \textbf{E}=-\frac{1}{c}\partial_{t}\textbf{B},\end{equation}
\begin{equation}\label{RHD2021ClLM div E with time} \nabla \cdot\textbf{E}=4\pi(en_{i}-en_{e}),\end{equation}
and
\begin{equation}\label{RHD2021ClLM rot B with time}
\nabla\times \textbf{B}=\frac{1}{c}\partial_{t}\textbf{E}+\frac{4\pi q_{e}}{c}n_{e}\textbf{v}_{e},\end{equation}
where the ions exist as the motionless background.

The Maxwell equations show that the source of electromagnetic field is
the density of the charge (presented via the concentration of particles)
and the density of the charge current expressed via the concentration of particles and velocity field.
Therefore, the hydrodynamic equations should have equation of the velocity field evolution.
However, majority of relativistic hydrodynamic models are based on the momentum density evolution equation \cite{Berezhiani PRL 94}.
The momentum density can be easily expressed via the velocity field for the cold plasmas,
like relativistic monoenergetic beams,
but there is no simple expression for the thermally distributed plasmas.
Nevertheless, existing hydrodynamic models contains equation of state for the momentum density.
But it is questionable, why do we need to consider a physical parameter
which is dropped at the first step.
Moreover, this transition is made in the first term of the equation in some approximate form.
All of it give some motivation for the derivation of the novel structure of the relativistic model presented above.
This model gives the straightforward derivation of the velocity field evolution equation with no relation to other physical parameters.


\section{Waves in the relativistic magnetized plasmas}

\subsection{Equilibrium state and the linearized hydrodynamic equations}

We consider small amplitude collective excitations relatively the macroscopically motionless equilibrium state of the relativistically hot
plasmas with anisotropic second rank tensors $p^{ab}$ and $t^{ab}$
(which are pressure like tensors while the pressure itself is not included in the model).
This equilibrium state is described by the relativistic Maxwellian distribution.
The equilibrium state is described within equilibrium concentration $n_{0}$ and temperature $T$.
The velocity field $\textbf{v}_{0}$ in the equilibrium state is equal to zero.
The equilibrium electric field $\textbf{E}_{0}$ is equal to zero.
The plasma is located in the constant and uniform external magnetic field $\textbf{B}_{0}=B_{0}\textbf{e}_{z}$
which create anisotropy in the system revealing in the anisotropy of tensors $p^{ab}$, $t^{ab}$, and $M^{abcd}$.
Two second rank tensors and one fourth rank tensor are involved in the description of the thermal effects.
Their structure is demonstrated after equation (\ref{RHD2021ClLM Euler for v}) for tensors $p^{ab}$ and $t^{ab}$,
\textit{and} equation (\ref{RHD2021ClLM eq for t a}) for tensor $M^{abcd}$.

We consider propagation of perturbations in the direction parallel to the external magnetic field $\textbf{k}=\{ 0,0, k_{z}\}$.
Let us present the linearized equations for the plane wave excitations propagating parallel to the external magnetic field.
The concentration appears as $\delta n=N_{0}e^{-\imath\omega t+\imath k_{z}z}$,
where $\omega$ is the frequency, $N_{0}$ is the constant amplitude.
Perturbations of other functions have same structure.
We start with the linearized continuity equation (\ref{RHD2021ClLM cont via v}):
\begin{equation}\label{RHD2021ClLM continuity equation lin 1D}
-\imath\omega\delta n+n_{0}\imath k_{z} \delta v_{z}=0. \end{equation}
Next, we show the linearized equations for the evolution of the three projections of velocity field obtained
from equation (\ref{RHD2021ClLM Euler for v})
\begin{equation}\label{RHD2021ClLM velocity field evolution equation lin 1D x}
-\imath\omega n_{0}\delta v_{x}
=\frac{q_{e}}{m}\biggl(\Gamma_{0} -\frac{\tilde{t}_{0}}{c^{2}}\biggr)\delta E_{x} +\Omega_{e}(\Gamma_{0}\delta v_{y}+\delta t_{y}),
\end{equation}
\begin{equation}\label{RHD2021ClLM velocity field evolution equation lin 1D y}
-\imath\omega n_{0}\delta v_{y}
=\frac{q_{e}}{m}\biggl(\Gamma_{0} -\frac{\tilde{t}_{0}}{c^{2}}\biggr)\delta E_{y}-\Omega_{e}(\Gamma_{0}\delta v_{x}+\delta t_{x}),
\end{equation}
and
\begin{equation}\label{RHD2021ClLM velocity field evolution equation lin 1D z}
-\imath\omega n_{0}\delta v_{z} +\imath k_{z}\frac{\delta p_{\parallel}}{m}
=\frac{q_{e}}{m}\Gamma_{0} \delta E_{z}-\frac{q_{e}}{mc^{2}}\tilde{t}_{0}\delta E_{z},
\end{equation}
where
$\Omega_{e}=q_{e}B_{0}/mc$ is the cyclotron frequency.

Equations (\ref{RHD2021ClLM continuity equation lin 1D}),
(\ref{RHD2021ClLM velocity field evolution equation lin 1D x}),
(\ref{RHD2021ClLM velocity field evolution equation lin 1D y}),
and
(\ref{RHD2021ClLM velocity field evolution equation lin 1D z})
require contribution of the flux of the average reverse gamma factor.
So, we use the linearized form of equation (\ref{RHD2021ClLM eq for t a}):
$$-\imath\omega \delta t_{x} -\imath\omega \Gamma_{0}\delta v_{x}$$
$$=\biggl(\frac{q}{m}n_{0}\delta E_{x} +\Omega_{e}n_{0}\delta v_{y}\biggr)\biggl(1-\frac{4p_{0,\perp}+p_{0,\parallel}}{n_{0}c^{2}}\biggr)$$
\begin{equation}\label{RHD2021ClLM eq for t x lin} +\frac{2q}{mc^{4}}(4M_{\perp} +M_{\parallel,\perp})\delta E_{x}, \end{equation}
and
$$-\imath\omega \delta t_{y} -\imath\omega \Gamma_{0}\delta v_{y}$$
$$=\biggl(\frac{q}{m}n_{0}\delta E_{y} -\Omega_{e}n_{0}\delta v_{x}\biggr)\biggl(1-\frac{4p_{0,\perp}+p_{0,\parallel}}{n_{0}c^{2}}\biggr)$$
\begin{equation}\label{RHD2021ClLM eq for t y lin} +\frac{2q}{mc^{4}}(4M_{\perp} +M_{\parallel,\perp})\delta E_{y}. \end{equation}

The Maxwell equations in the linear approximation for waves propagating parallel to the external magnetic field have well-known form
\begin{equation}\label{RHD2021ClLM Maxwell lin wave x}
(\omega^{2}-k_{z}^{2}c^{2})\delta E_{x}+4\pi q_{e}\imath\omega n_{0}\delta v_{x}=0, \end{equation}
\begin{equation}\label{RHD2021ClLM Maxwell lin wave y}
(\omega^{2}-k_{z}^{2}c^{2})\delta E_{y}+4\pi q_{e}\imath\omega n_{0}\delta v_{y}=0, \end{equation}
and
\begin{equation}\label{RHD2021ClLM Maxwell lin wave z}
\omega^{2}\delta E_{z}+4\pi q_{e}\imath\omega n_{0}\delta v_{z}=0. \end{equation}

\subsection{Spectra of waves}

Equations (\ref{RHD2021ClLM continuity equation lin 1D}),
(\ref{RHD2021ClLM velocity field evolution equation lin 1D z}),
(\ref{RHD2021ClLM Maxwell lin wave z})
described the longitudinal waves.
They lead to the following spectrum of the relativistic Langmuir wave
\begin{equation}\label{RHD2021ClLM spectrum Langmuir}
\omega^{2}=\biggl(\frac{\Gamma_{0}}{n_{0}}-\frac{t_{0,\parallel}}{n_{0}c^{2}}\biggr)\omega^{2}_{Le}
+\frac{\partial p_{0,\parallel}}{\partial n_{0}}k_{z}^{2},
\end{equation}
where it is assumed that the flux of the particle current $p_{0,\parallel}$ is the functional of the concentration
$\delta p_{\parallel}= (\partial p_{0,\parallel}/\partial n_{0})\delta n$.

Let us to look on the high-frequency transverse waves propagating parallel to the magnetic field.
They are described within equations
(\ref{RHD2021ClLM velocity field evolution equation lin 1D x}),
(\ref{RHD2021ClLM velocity field evolution equation lin 1D y}),
(\ref{RHD2021ClLM eq for t x lin}),
(\ref{RHD2021ClLM eq for t y lin}),
(\ref{RHD2021ClLM Maxwell lin wave x}),
and
(\ref{RHD2021ClLM Maxwell lin wave y}).
The Maxwell equations (\ref{RHD2021ClLM Maxwell lin wave x}),
and
(\ref{RHD2021ClLM Maxwell lin wave y}) contain the velocity perturbations.
Therefore, we need to extract the velocity perturbations from equations
(\ref{RHD2021ClLM velocity field evolution equation lin 1D x}),
(\ref{RHD2021ClLM velocity field evolution equation lin 1D y}),
(\ref{RHD2021ClLM eq for t x lin}),
and
(\ref{RHD2021ClLM eq for t y lin}).
They appear in the following form
$$\delta v_{x}=\frac{\imath\omega}{\omega^{2}-f_{0}\Omega^{2}}
\biggl[\frac{q}{m}\delta E_{x}\biggl(\frac{\Gamma_{0}}{n_{0}}-\frac{t_{0,\perp}}{n_{0}c^{2}}\biggr)$$
\begin{equation}\label{RHD2021ClLM vel x expr via E}
+\frac{q}{m}f_{0}\frac{\imath \Omega}{\omega}\delta E_{y}
+\frac{2q}{c}\frac{f_{0M}}{n_{0}}\frac{\imath \Omega}{\omega}\delta E_{y}\biggr],
\end{equation}
and
$$\delta v_{y}=\frac{\imath\omega}{\omega^{2}-f_{0}\Omega^{2}}
\biggl[\frac{q}{m}\delta E_{y}\biggl(\frac{\Gamma_{0}}{n_{0}}-\frac{t_{0,\perp}}{n_{0}c^{2}}\biggr)$$

\begin{equation}\label{RHD2021ClLM vel y expr via E}
-\frac{q}{m}f_{0}\frac{\imath \Omega}{\omega}\delta E_{x}
-\frac{2q}{c}\frac{f_{0M}}{n_{0}}\frac{\imath \Omega}{\omega}\delta E_{x}\biggr],
\end{equation}
where we use the following notations
$f_{0}=1-(4p_{0,\perp}+p_{0,\parallel})/(n_{0}c^{2})$,
and
$f_{0M}=(4M_{0,\perp}/3 +M_{0,\parallel,\perp})/(mc^{2})$.

Let us write down the dispersion equation as the determinant of the corresponding part of the dielectric permeability tensor:
\begin{widetext}
\begin{equation}\label{RHD2021ClLM det to spectrum tr}
\Biggl|\begin{array}{cc}
\omega^{2}-k_{z}^{2}c^{2}-\frac{\omega^{2}_{Le} \omega^{2}}{\omega^{2}-f_{0} \Omega^{2}}
\biggl(\frac{\Gamma_{0}}{n_{0}}-\frac{t_{0,\perp}}{n_{0}c^{2}}\biggr) &
-\frac{\imath\omega \Omega \omega^{2}_{Le}}{\omega^{2}-f_{0} \Omega^{2}}\biggl(f_{0}+\frac{2mf_{0M}}{n_{0}c^{2}}\biggr) \\
\frac{\imath\omega \Omega \omega^{2}_{Le}}{\omega^{2}-f_{0} \Omega^{2}}\biggl(f_{0}+\frac{2mf_{0M}}{n_{0}c^{2}}\biggr) &
\omega^{2}-k_{z}^{2}c^{2}-\frac{\omega^{2}_{Le} \omega^{2}}{\omega^{2}-f_{0} \Omega^{2}}
\biggl(\frac{\Gamma_{0}}{n_{0}}-\frac{t_{0,\perp}}{n_{0}c^{2}}\biggr)
\end{array}
\Biggr|=0.
\end{equation}
\end{widetext}

The cut-off frequency of the longitudinal wave (\ref{RHD2021ClLM spectrum Langmuir}) is bound to $t_{0,\parallel}$
which defines the coefficient in front of the Langmuir frequency.
However, the cut-off frequencies for the transverse waves contains another coefficient in front of the Langmuir frequency
which is proportional to $\Gamma_{0}/n_{0}-t_{0,\perp}/n_{0}c^{2}$.
It includes another element of tensor $\hat{t}$: $t_{0,\perp}$.
The group velocity of the relativistic Langmuir wave (\ref{RHD2021ClLM spectrum Langmuir}) is proportional to derivative of $p_{0,\parallel}$.
If we consider the electromagnetic wave propagation perpendicular to the magnetic field with the linear polarization of electric field changing parallel to the magnetic field
we find the following spectrum
\begin{equation}\label{RHD2021ClLM spectrum EM with k perp B}
\omega^{2}=\biggl(\frac{\Gamma_{0}}{n_{0}}-\frac{t_{0,\parallel}}{n_{0}c^{2}}\biggr)\omega^{2}_{Le}
+k_{x}^{2}c^{2}.
\end{equation}
This result underline the various contribution of $t_{0,\parallel}$, $t_{0,\perp}$, and $p_{0,\parallel}$.
It has same cut-off frequency as the longitudinal wave (\ref{RHD2021ClLM spectrum Langmuir}) propagating parallel to the magnetic field.
But they obviously have different group velocities.

\section{Conclusion}

General derivation of the hydrodynamic model for the relativistically hot plasmas has been presented in earlier papers
\cite{Andreev 2021 05}, \cite{Andreev 2021 09}.
This hydrodynamic model is based on the dynamics of
four material fields: the concentration and the velocity field
\emph{and} the average reverse relativistic $\gamma$ factor and the flux of the reverse relativistic $\gamma$ factor.
However, the final truncation has been made for the isotropic structure of the high-rank tensors entering the model.
In this paper, we have presented the generalization of the model including the anisotropy of the high-rank tensors.
It has been assumed that there is the single anisotropy direction related to the external magnetic field.
It leads to two values of the flux of current of particles $\hat{p}$
and two values of the current of the flux of reverse gamma factor $\hat{t}$.
The fourth rank tensor $\hat{M}$ has three independent elements.
Experimental detection of the anisotropic effects can be partially made through the analysis of spectrum of waves
including the measurement of the value of cut-off frequencies
and measurement of the group-velocities of waves.
Complete theoretical treatment of this problem requires further analysis of the anisotropic relativistic distribution function
which can give background for the calculation of the approximate calculation of equations of state.

The background for the further analysis of the linear and nonlinear wave phenomena
in the relativistically hot strongly magnetized anisotropic plasmas has been developed.
Moreover, it includes the anisotropy of the tensors describing the thermal effects.

\section{Acknowledgements}

Work is supported by the Russian Foundation for Basic Research (grant no. 20-02-00476).

\section{DATA AVAILABILITY}

Data sharing is not applicable to this article as no new data were
created or analyzed in this study, which is a purely theoretical one.

\end{document}